\documentclass[english]{revtex4-2}
\usepackage{babel}
\usepackage{amssymb}
\usepackage{graphicx}
\usepackage{color}
\usepackage{bm}
\usepackage{longtable}
\usepackage{amsmath} 
\usepackage{amsfonts}
\usepackage{amssymb}
\usepackage{mathtools}
\usepackage[T1]{fontenc}
\usepackage[latin1]{inputenc}
\usepackage{braket}
\usepackage{babel}%
\usepackage{graphicx}
\usepackage{color}
\usepackage{bm}
\usepackage{longtable}
\usepackage{amsmath}
\usepackage{amsfonts}
\usepackage{dsfont}
\usepackage{amssymb}
\usepackage{hyperref}
\newcommand{\ba}{\begin{eqnarray}}
\newcommand{\ea}{\end{eqnarray}}

\newcommand{\etal}{{\it{et. al. }}}

\begin{document}
\title{Sharing nonlocality in a network using the quantum violation of chain network inequality}
\author{Rahul Kumar}
\affiliation{National Institute of Technology Patna, Ashok Rajpath, Patna, Bihar 800005, India}
\author{ A. K. Pan }
\affiliation{Department of Physics, Indian Institute of Technology Hyderabad, Telengana 502284, India }
\begin{abstract}
 Based on the quantum violation of suitable $n$-local inequality in a star network for arbitrary $m$ inputs, we demonstrate the sharing of nonlocality in the network. Such a network features an arbitrary $n$ number of independent sources, $n$ edge parties, and a central party. Each party receives arbitrary $m$ inputs. We consider two different types of sharing of nonlocality in the network. i)  The symmetric case - when the sharing of nonlocality is considered across all edge parties. ii) The asymmetric case -  when the sharing of nonlocality is considered across only one edge party. For simplicity, we first consider the bilocal scenario $(n=2)$ with three inputs $m=3$ and demonstrate that while in the symmetric case at most two sequential observers can share nonlocality, in the asymmetric case at most four sequential observers can share nonlocality. We extend the study to $n$-local scenario by assuming each party receives three inputs and show that in the symmetric case the result remains the same for any $n$, but in the asymmetrical case, an unbounded number of sequential observers can share nonlocality across one edge for a sufficiently large value of $n$. We further extend our result for arbitrary $m$ input in $n$-local scenario. We demonstrate that for $m\geq 4$, in the symmetric case at most one sequential observer can share nonlocality irrespective of the value of $n$. For the asymmetric case, we analytically show that there exists $n(k)$ for which an arbitrary $k$ number of sequential observers can share the nonlocality across one edge. The optimal quantum violation of $m$-input $n$-local inequality is derived through an elegant SOS approach without specifying the dimension of the quantum system. 
\end{abstract}
\maketitle
Bell's theorem is the most profound result to demonstrate the distinctive feature of quantum correlation over its classical counterpart \cite{bell,brun2014}. It states that any classical model that satisfies local realism cannot reproduce all statistics of quantum theory.  In a typical bipartite Bell experiment scenario, there are two distant parties Alice and Bob share a physical system originating from a single source. Alice (Bob) receives input $x$ ($y$) and her (his) measurement produces output $a$ ( $b$). In a  local realist model where one fixes the outcomes by a hidden variable $\lambda$ and outcomes are independent of any kind of distance influences, the joint probability of this experiment can be written in a factorized form as
\begin{equation}P(a,b|x,y) =  \int \rho (\lambda)\hspace{1mm}P(a|x,\lambda)\hspace{1mm} P(b|y,\lambda)\hspace{1mm} d\lambda\end{equation}
where $\rho(\lambda)$ is the distribution of $\lambda$ satisfying $\int\rho(\lambda) d\lambda=1$. 
In quantum theory, if the source distributes a suitable entangled state and the two parties perform locally incompatible measurements, the joint probability cannot always be factorized.`This distinctive feature is known as quantum nonlocality and is demonstrated through the quantum violation of various forms of Bell's inequalities \cite{brun2014}. The multipartite Bell experiment is a generalization of the bipartite scenario where multiple parties receive the physical system from a single source.  

Of late, there is an upsurge of interest in exploring the quantum nonlocality in networks. This multipartite nonlocality in a network is conceptually different from standard multipartite Bell nonlocality. While the latter case features only a single common source, in the former case  there are several independent sources. Branciard \etal
{\cite{bran2012}} firstly introduced the simplest non-trivial network scenario (well known bilocal scenario) involving three parties and two independent sources. Violation of suitable formulated nonlinear inequality {\cite{bran2012,bran2010}} shows quantum correlations in network. It is expanded to $n$-locality scenario {\cite{tava2016,tava2014,andr2017}} featuring $n$ number of sources and edge parties. In \cite{munshi2021,munshi2022}, the $n$-locality inequalities for arbitrary input scenario and their optimal quantum violations have been studied without assuming the dimension of the system. Recently network nonlocality has been studied in various topologies \cite{{Alejandro 2019},{Cyril Branciard 2012},{Renou 2019 },{Denis 2019},{Ming 2018},{tava2016},{rafael},{Tavakoli 2020},{Ivan 2020},{Nicolas 2017},{Tamas 2020},{Benjamin2021},{Xavier2021},{Patricia2021},{Bancal2021}, tava2014, Tavakoli 2017,andr2017,kundu,munshi2021,munshi2022,rahu2022}.

In this work, we study the sharing of network nonlocality by multiple independent sequential observers. In recent times, there has been an upsurge of interest in sharing various forms of quantum correlations  \cite{silva2014,Colbeck2020,Anwer2020,asmita,Debarshi,Shashank,Akshata,Shounak,saptarshi,Karthik,miklin,sumit,shyam2022,Cheng2022,cheng2021,hou2022,Ren2022,Zhang2021,Feng2020,Ren2019,Mao2022}. 
 Based on Clauser-Horne-Shimony-Holt (CHSH) inequality, Silva \emph{et al.} \cite{silva2014} first demonstrated the sharing of quantum nonlocality by two sequential observers for one party which was later extended \cite{Colbeck2020} to arbitrary sequential observers.   Inspired by Silva \emph{et al.} \cite{silva2014}, quite a number of studies have been made to demonstrate the sharing of various forms of quantum correlations, such as, preparation contextuality \cite{asmita}, steering \cite{Shashank,Akshata}, and entanglement \cite{saptarshi}. Sharing the quantum advantage in the prepare-measure communication game and self-testing of unsharpness parameter has also demonstrated  \cite{Karthik,sumit,miklin}.  Recently, the recycling of nonlocal resources in a quantum network has been briefly studied in \cite{Mao2022,Cheng2022,shyam2022}. In \cite{Mao2022}, the authors demonstrate that the nonlocality of a star network can only be shared by the first edge parties, not by secondary edge parties. In \cite{shyam2022}, one of us demonstrated the sharing of nonlocality by an unbounded number of observers in one edge of the arbitrary input star-network scenario. It is also shown \cite{cheng2021} that sharing cannot be demonstrated by two sequential observers at both ends.

In this paper, we consider a star network scenario that features arbitrary $n$ independent sources, $n$ edge observers, and a central observer. Each edge observer shares an independent physical system with the central party, originating from an independent source. We introduce the network chain inequality featuring arbitrary $m$ inputs per edge party. We demonstrate the sequential sharing of network nonlocality by an unbounded number of independent observers across one edge of the star network. The scenario is different compared to \cite{shyam2022} where a different network inequality was considered. The quantum violation of network chain inequality can be optimized for a two-qubit entangled state shared between each edge party and the central party, compared to the inequality used in \cite{shyam2022} that requires a higher dimensional system. 

We first introduce the bilocal scenario $(n=2)$ where each of the three parties receives three inputs $(m=3)$. We propose a bilocality inequality and demonstrate the optimal quantum violation using an elegant method named the sum-of-squares (SOS) approach. Such an optimization technique fixes the observables and the joint state of the system without requiring to mention the dimension of the system. Further, we show the sharing of nonlocality in two different cases. i)  The symmetric case - when the sharing of nonlocality is considered for both the edge parties. We find that a maximum of two sequential observers across each edge can share the nonlocality. ii) The asymmetric case -  when the sharing of nonlocality is considered across one of the two edges. We find that a maximum of four sequential observers can share the nonlocality. We extend our study for arbitrary $n$ parties in a star network by taking three inputs for each edge party and propose a $n$-locality inequality. We show that in the symmetric case a maximum of two observers in each edge can share nonlocality as in the bilocal scenario. But in an asymmetric case, there is an increase in the number of sequential observers across one edge if the number of parties $n$ increases. An unbounded number of sequential observers can share nonlocality for a sufficiently large value of $n$. 
	
We further extend our study for arbitrary $m$ inputs in $n$-locality scenario and propose chain $n$-locality inequality. We find that for $m\geq 4$, at most one sequential observer can share nonlocality for an arbitrary number of $n$ in the symmetric case. But there is a sharp increase in the number of the sequential observers which can exhibit sharing of nonlocality across one edge with an increase in value of $n$. Finally, we provide a generalized analytical relation valid for any arbitrary input $m$ between the number of sequential observers $k$ that can share nonlocality and the number of edge parties $n$.  We note again that throughout this work, the dimension of the system remains unspecified.

The paper is organized as follows. In Sec. I,  we introduce chain bilocality inequality for three inputs ($m=3$) scenarios and derive its optimal quantum violation using the SOS approach without assuming the dimension of the system.   In Sec. II,  we demonstrate the sharing of nonlocality for three input scenarios for the symmetric and asymmetric cases. In Sec. III, we extend the sharing of nonlocality in $n-$local scenario for three inputs.  In Sec. IV, we generalize the sharing of nonlocality in $n-$locality scenario for arbitrary $m$ input in both cases and derive its optimal quantum violation of chain $n$-locality inequality without assuming the dimension of the system.

\section{ BILOCAL NETWORK FOR three inputs ($m=3$)}
\begin{figure}[ht]
\centering
{{\includegraphics[width=1.0\linewidth]{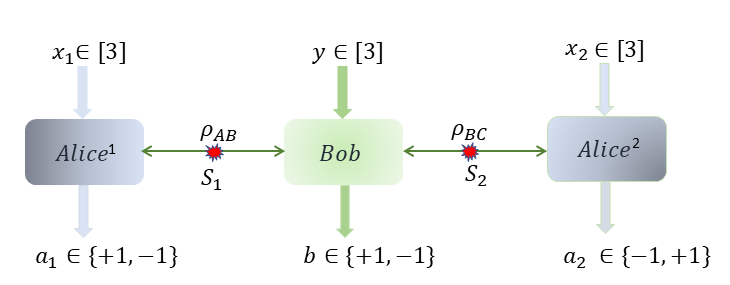}}}
\caption{Bilocal scenario for three inputs}
\label{fig1}
\end{figure}

We start with bilocal scenario \cite{bran2012} that features three inputs, i.e., $m=3$. As depicted in Fig. (\ref{fig1}), there are two independent sources $S_1$ and $S_2$, and total of three parties Alice$^1$, Alice$^2$, and Bob respectively. The source $S_{1}$ sends physical systems to Alice$^1$ and $Bob$ and the source $S_{2}$ sends physical systems to Bob and Alice$^2$. In the bilocal scenario $(n=2)$, Alice$^1$ (Alice$^2$) performs three dichotomic measurements $A^1_{1}$, $A^1_{2}$ and $A^1_{3}$ ($A^2_{1}$, $A^2_{2}$and $A^2_3$) corresponding to the input $x_1\in [3]$ ($x_2\in[3]$), and Bob also performs three measurement $B_{1}$, $B_{2}$ and $B_{3}$  corresponding to the input $y\in[3]$ and produce outputs $a_1, b, a_2\in \{+1,-1\}$ respectively. 

In particular, Bob performs measurement on the joint physical systems he receives from two sources $S_{1}$ and $S_{2}$ which are assumed to be independent of each other - the bilocality assumption. If two sources $S_{1}$ and $S_{2}$ produce physical systems  $\lambda_{1}$ and $\lambda_{2}$ having distribution $\rho(\lambda_{1},\lambda_{2})$, then according to bilocality assumption  $\rho{(\lambda_{1},\lambda_{2})}$ = $\rho_{1}{(\lambda_{1})}$ $\rho_{2}{(\lambda_{2})}$, and $\rho_{1}{(\lambda_{1})}$ and $\rho_{2}{(\lambda_{2})}$ are independent distribution that satisfy  $\int d\lambda_{1}\rho_{1}{(\lambda_{1})}=1$ and $\int d\lambda_{2}\rho_{2}{(\lambda_{2})}=1$.  The joint probability can be written as
\begin{eqnarray}
\nonumber
P(a_1, b, a_2|x_1,y,x_2)&=&\int\int d\lambda_{1} d\lambda_{2}\hspace{2mm}\rho_{1}(\lambda_{1})\hspace{1mm}\rho_{2}(\lambda_{2})\\
&&\times P(a_1|x_1,\lambda_{1})P(b|y,\lambda_{1},\lambda_{2})P(a_2|x_2,\lambda_{3}).\hspace{9pt}
\end{eqnarray}
The measurement outcome  of Alice$^1$ (Alice$^2$)   solely depend on $\lambda_{1}$($\lambda_{2}$) but outcome of Bob depends on both $\lambda_{1}$ and $\lambda_{2}$.  

 We define a suitable linear combination of correlations are given as;
\begin{eqnarray}
\label{m=3}
\nonumber
&&J^2_{3,1}=\hspace{1mm}{\langle}(A^1_{1}+A^1_{2})B_{1}(A^2_{1}+A^2_{2}){\rangle},\\
&& J^2_{3,2}=\hspace{1mm}{\langle}(A^1_{2}+A^1_{3})B_{2}(A^2_{2}+A^2_{3}){\rangle},\\
\nonumber
&&J^2_{3,3}=\hspace{1mm}{\langle}(A^1_{3}-A^1_{1})B_{3}(A^2_{3}-A^2_{1}){\rangle}
\end{eqnarray}
Here superscript 2 in $J^2_{3,1}$ denotes bilocality scenario, i.e., $n=2$ and subscript 3 denotes number of inputs given, i.e., $m=3$.\\
 We propose the following chain bilocality inequality 
\begin{equation}
\label{beta23i}
(\beta^2_{3})_{bl}=\sqrt{|{J}^2_{3,1}|}+\sqrt{|{J}^2_{3,2}|}  + \sqrt{|{J}^2_{3,3}|}\leq 4
\end{equation}

The above bound can be derived by using bilocality assumption so that  defining $\langle{A_{x_1}}\rangle_{\lambda_{1}} = \sum_{a_1}(-1)^{a_1}  P(a_1|x_1)$ and $\langle{A_{x_2}}\rangle_{\lambda_{2}} = \sum_{a_2}(-1)^{a_2}  P(a_2|x_2)$ where $x_1,x_2\in [3]$  along with considering $|\langle{B_{1}}\rangle_{\lambda_{1},\lambda_{2}}|\leq{1},$ we can write
\begin{eqnarray}
	\label{Beta1}
	|J^2_{3,1}|&\leq&\int{d}\lambda_{1}\rho_{1}(\lambda_{1})|\langle{A^1_{1}}\rangle_{\lambda_{1}}+\langle{A^1_{2}}\rangle_{\lambda_{1}}|\\
	\nonumber
	&\times&
	\int{d}\lambda_{3}\rho_{3}(\lambda_{3})|\langle{A^2_{1}}\rangle_{\lambda_{3}}+\langle{A^2_{2}}\rangle_{\lambda_{3}}|. \hspace{5mm}
\end{eqnarray}
The terms $|J^2_{3,2}|$ and $|J^2_{3,3}|$ given in  Eq. (\ref{m=3}) can also be written in similar manner. Using the inequality
\begin{equation}
	\label{Beta7}
	\sum\limits_{i=1}^{n}\sqrt{r_{i}s_{i}}\leq\sqrt{\sum\limits_{i=1}^{n}}r_{i}\sqrt{\sum\limits_{i=1}^{n}s_{i}}
\end{equation} 
for $r_{i}, s_{i} \geq 0$, $i\in[3]$, we can write from Eq.(\ref{beta23i}) that 
\begin{equation}
	(\beta^2_3)_{bl}\leq \sqrt{\int{d}\lambda_{1}\rho_{1}(\lambda_{1})\eta^1_A }\times \sqrt{\int{d}\lambda_{3}\rho_{3}(\lambda_{3})\eta^2_A },
\end{equation}
where
\begin{eqnarray} 
\eta_A^1 = |{\langle A^1_1\rangle}_{\lambda_{1}}+{\langle A^1_2\rangle}_{\lambda_{1}}|+|{\langle A^1_2\rangle}_{\lambda_{1}}+|{\langle A^1_3\rangle}_{\lambda_{1}}|+|{\langle A^1_3\rangle}_{\lambda_{1}}-|{\langle A^1_1\rangle}_{\lambda_{1}}|\\
\eta_A^2= |{\langle A^2_1\rangle}_{\lambda_{3}}+{\langle A^2_2\rangle}_{\lambda_{3}}|+|{\langle A^2_2\rangle}_{\lambda_{3}}+|{\langle A^2_3\rangle}_{\lambda_{3}}|+|{\langle A^2_3\rangle}_{\lambda_{3}}-|{\langle A^2_1\rangle}_{\lambda_{3}}|
\end{eqnarray}

Since all the observables are dichotomic having values $\pm1$, it is simple to check that $\eta^1_A$=$\eta^2_A\leq 4$. Integrating over $\lambda_{1}$ and $\lambda_{3}$ we obtain $(\beta^2_3)_{bl}\leq 4$, as claimed in Eq. (\ref{beta23i}).
\subsection{Optimal violation of chain bell inequality for $m=3$ in bilocal scenario}
\label{SECVA}

To optimize the quantum value of $(\beta^2_{3})_Q$ in  Eq. (\ref{beta23i}), we follow an elegant SOS approach.   We define a positive semidefinite operator $\langle \gamma^2_{3}\rangle_{Q}\geq 0$ so that  $\langle \gamma^2_{3}\rangle_Q=-\tau^2_{3}+(\beta^2_{3})_Q$ where $\tau^2_{3}$ is the optimal quantum value of $(\beta^2_{3})_Q$. Here again superscript 2 and subscript 3 in $\gamma^2_{3}$, $\tau^2_{3}$ and $\beta^2_{3}$ denotes bilocal scenario, i.e., $n=2$ and three number of inputs i.e., $m=3$ respectively.	By considering suitable positive operators $M^2_{3,i}$, where $ i\in[3]$ we can write
	\begin{equation}
	\label{gammach3}
	\gamma^2_{3}=\sum_{i=1}^{3}\dfrac{\sqrt{\omega^2_{3,i}}}{2}{|M^2_{3,i}|\psi\rangle|}^2
	\end{equation} 
 where $\omega^2_{3,i}$s are suitable positive numbers and $\omega^2_{3,i}$=$\omega^{A_1}_{3,i}$ $\cdot$ $\omega^{A_2}_{3,i}$ that will be specified soon. Here $\ket{\psi}_{A_1BA_2} = \ket{\psi}_{A_1B}\otimes\ket{\psi}_{BA_2}$ are originating from independent sources $S_{1}$ and $S_{2}$ respectively. For notational convenience we denote $\ket{\psi}_{A_1BA_2} = \ket{\psi}$. For our purpose, we choose  $M^2_{3,i}$ where $i\in[3]$ as
	
\begin{eqnarray} 
\label{meas3}
\nonumber
&&|M^2_{3,1}|\psi\rangle|=\sqrt{\bigg|\left(\frac{A^1_{1}+A^1_{2}}{\omega^{A_1}_{3,1}}\otimes\frac{A^2_{1}+A^2_{2}}{(\omega^{A_2}_{3,1})}\right)|\psi\rangle\bigg|} -\sqrt{| B_{1}|\psi\rangle|}\\
\nonumber
&&|M^2_{3,2}|\psi\rangle|=\sqrt{\bigg|\left(\frac{A^1_{2}+A^1_{3}}{\omega^{A_1}_{3,2}}\otimes\frac{A^2_{2}+A^2_{3}}{(\omega^{A_2}_{3,2})}\right)|\psi\rangle\bigg|} -\sqrt{| B_{2}|\psi\rangle|}\\
\nonumber
&&|M^2_{3,3}|\psi\rangle|=\sqrt{\bigg|\left(\frac{A^1_{3}-A^1_{1}}{\omega^{A_1}_{3,3}}\otimes\frac{A^2_{3}-A^2_{1}}{(\omega^{A_2}_{3,3})}\right)|\psi\rangle\bigg|} -\sqrt{| B_{3}|\psi\rangle|}\\
\end{eqnarray}	
where
\begin{eqnarray}
\label{omega3}
\nonumber
&&\omega^{A_1}_{3,1}=||(A^1_{1}+A^1_{2})\ket{\psi}||_2 \ ; \ \ \omega^{A_2}_{3,1}=||(A^2_{1}+A^2_{2})\ket{\psi}||_2 \\
\nonumber
&&\omega^{A_1}_{3,2}=||(A^1_{2}+A^1_{3})\ket{\psi}||_2 \ ; \ \ \omega^{A_2}_{3,2}=||(A^2_{2}+A^2_{3})\ket{\psi}||_2 \\
\nonumber
&&\omega^{A_1}_{3,3}=||(A^1_{3}-A^1_{1})\ket{\psi}||_2 \ ; \ \ \omega^{A_2}_{3,3}=||(A^2_{3}-A^2_{1})\ket{\psi}||_2 \\
\end{eqnarray}
Putting $M^2_{3,i}$ from Eq.~(\ref{meas3}) into the Eq. (\ref{gammach3}), after a simple algebraic evaluation obtain
\begin{equation} 
 \label{sos2}
\langle \gamma^2_3 \rangle =\sum\limits_{i=1}^3 \sqrt{\omega^2_{3,i}}-(\beta^2_3)_Q
\end{equation}
Clearly, it follows that the quantum optimal value corresponds $\langle \gamma^2_3 \rangle=0$. Therefore,
\begin{eqnarray}
 \label{SQopt3}
\left(\beta^2_{3}\right)_{Q}^{opt} = max\left(\sum\limits_{i=1}^3 \sqrt{\omega^{A_1}_{3,i}\cdot\omega^{A_2}_{3,i}}\right)
\end{eqnarray}
	
	Note that, the optimal quantum value will occur when $\langle\gamma^2_{3}\rangle=0$, which in turn gives the optimisation condition as follows
\begin{equation}
 \label{impn3s1}
|M^2_{3,i}\ket{\psi}|=0 \ \implies M^2_{3,i}\ket{\psi}=0 \ \ \forall i \in \{1,2,3\}
\end{equation}
	
	Now, in order to evaluate the optimum quantum value $\left(\beta^2_{3}\right)_{Q}^{opt}$ and thus the quantity $\sum\limits_{i=1}^{3}\sqrt{\omega_{3,i}^{A_1}\cdot \omega_{3,i}^{A_2}}$. By using the inequality
 \begin{equation}
\label{Tavakoli}
\ \forall\ \ z_{i}^{k} \geq 0; \ \ \ \sum\limits_{i=1}^{m}\bigg(\prod\limits_{k=1}^{n}z_{i}^{k}\bigg)^{\frac{1}{n}}\leq \prod \limits_{k=1}^{n}\bigg(\sum\limits_{i=1}^{m}z_{i}^{k}\bigg)^{\frac{1}{n}}
\end{equation}
 the right-hand-side of Eq. (\ref{SQopt3}) reduces to the following 
\begin{equation}
\label{omega1s1}
\sum\limits_{i=1}^{3}\left(\prod \limits_{k=A_1,A_2}\omega^{k}_{3,i}\right)^{\frac{1}{2}}\leq\prod \limits_{k=A_1,A_2}\left(\sum\limits_{i=1}^{3}\omega^{k}_{3,i}\right)^{\frac{1}{2}}	
\end{equation}
	
	Further, by applying the convex inequality
 \footnote{\label{f3}
 From the Jensen's inequality given by $f(\sum\limits_{k=1}^{n}r_k x_k) \leq \sum\limits_{k=1}^{n} r_k f(x_k) $ where $\sum\limits_{k=1}^{n}r_k =1$, the following inequality can be derived 
 \begin{equation} 
 \sum\limits_{k=1}^{n}\omega_{k}\leq \sqrt{\ n\sum\limits_{k=1}^{n}\omega^2_k} 
 \end{equation}}
 the quantity $\sum\limits_{j=1}^{3}\omega^{k}_{3,i}$  can be written as
	\begin{equation}
		\label{omega2s1}
		\sum\limits_{i=1}^{3}(\omega^{k}_{3,i})\leq \sqrt{3\sum\limits_{i=1}^{3}\bigg(\omega^k_{3,i}\bigg)^{2}}
	\end{equation}
	
	Then, by combining Eq. (\ref{omega1s1}) and (\ref{omega2s1}), from Eq. (\ref{SQopt3} ) we obtain 
	\begin{eqnarray}\label{s3m1}
		\left(\beta^2_{3}\right)_{Q}^{opt} \leq \max\left[\prod_{k=A_1,A_2}^{}\left(3\sum_{i=1}^{3}\left( \omega^{k}_{3,i}\right)^2\right) \right] ^{\frac{1}{4}}
	\end{eqnarray}
	
	where each $\left( \omega^{k}_{3,i}\right)^2$ from Eq. (\ref{omega3}) can be written as 
\begin{eqnarray} 
\label{omegaA1}
\nonumber
\left(\omega^{A_1}_{3,1}\right)^2&=&\bra{\psi}(2+\{A^1_{1},A^1_{2}\})\ket{\psi}\label{w311}\\
\nonumber
\left(\omega^{A_1}_{3,2}\right)^2&=&\bra{\psi}(2+\{A^1_{2},A^1_{3}\})\ket{\psi}\label{w322}\\
\left(\omega^{A_1}_{3,3}\right)^2&=&\bra{\psi}(2-\{A^1_{3},A^1_{1}\})\ket{\psi}\\
\label{omegaA2}
\nonumber
\left(\omega^{A_2}_{3,1}\right)^2&=&\bra{\psi}(2+\{A^2_{1},A^2_{2}\})\ket{\psi}\label{w311}\\
\nonumber
\left(\omega^{A_2}_{3,2}\right)^2&=&\bra{\psi}(2+\{A^2_{2},A^2_{3}\})\ket{\psi}\label{w322}\\
\nonumber
\left(\omega^{A_2}_{3,3}\right)^2&=&\bra{\psi}(2-\{A^2_{3},A^2_{1}\})\ket{\psi}\\
\label{w333}
\end{eqnarray}
   Now, in the following we calculate $\sum_{i=1}^{3}\left( \omega^{A_1}_{3,i}\right)^2$ from Eq. (\ref{omegaA1})  and  $\sum_{j=1}^{3}\left( \omega^{A_2}_{3,i}\right)^2$ from Eq. (\ref{omegaA2}) separately.\\
	
	\paragraph{\textit{Evaluation of $\sum_{i=1}^{3}\left( \omega^{A_1}_{3,i}\right)^2$:}}\label{pws1}
	\begin{equation}\label{v32}
		\sum_{i=1}^{3}\left( \omega^{A_1}_{3,i}\right)^2= \bra{\psi}(6+ \{A^1_{2},(A^1_{1}+A^1_{3})\}-\{A^1_{1},A^1_{3} \})\ket{\psi}\nonumber
		\end{equation}
   Considering $A^1_2=(A^1_1+A^1_3)/\omega'_3$, we get
 \begin{equation}
 \sum_{i=1}^{3}\left(\omega^{A_1}_{3,i}\right)^2=\bra{\psi}6+2\sqrt{2+\langle\{A^1_1,A^1_3)\}\rangle}-\langle\{A^1_1,A^1_3)\}\rangle\ket{\psi}
 \end{equation}
 A simple calculation gives the maximization condition  $\{A^1_1,A^1_3)\}=-1$ which implies $\omega'_3=\sqrt{2+\{A^1_1,A^1_3)\}}=1$. Thus,  we get the condition on Alice's observables  $A^1_1-A^1_2+A^1_3=0$. Also, we have found $\{A^1_1,A^1_2\}=1$ and $\{A^1_2,A^1_3\}=1$ and consequently  $\omega^{A^1}_{3,1}=\omega^{A^2}_{3,2}=\omega^{A^3}_{3,3}=\sqrt{3}$. Bob's observables and the state required for this optimization can also be found from the condition $\langle\gamma^2_3\rangle_Q=0$, i.e., $\forall i\ \ \ M^2_{3,i}\ket{\psi} =0 $.
	
	Similarly for $\sum_{i=1}^{3}\left( \omega^{A_2}_{3,i}\right)^2$, we get the same value. We then obtain the optimal quantum value by using the value of  Eq. (\ref{v32}) in Eq. (\ref{SQopt3}), we get
	\begin{eqnarray}\label{s3qm}
		\left(\beta^2_{3}\right)_{Q}^{opt}= 3\sqrt{3}	
	\end{eqnarray}
	
	It is important to remark here that the optimal quantum value $\left(\beta^2_{3}\right)_{Q}^{opt}= 3\sqrt3$ is evaluated without specifying the dimension of the system. The optimal value fixes the states, and the observables are the following. 
\subsection{State and observables for obtaining  $\left(\beta^2_{3}\right)_{Q}^{opt}$}	
	
	Through SOS we  established the relationships between the observables of all the parties to achieve the optimal quantum violation. The relationship is $A^1_1-A^1_2+A^1_3=0$, which in turn provides the relation between the observables in terms of the anti-commuting relations.
	
	\begin{eqnarray}
		&&\{A^1_{1},A^1_{2}\}=\{A^1_{2},A^1_{3}\}=-\{A^1_{1},A^1_{3}\}=\mathbb{I}_{d}\label{v43}\\
		&&\{A^2_{1},A^2_{2}\}=\{A^2_{2},A^2_{3}\}=-\{A^2_{1},A^2_{3}\}=\mathbb{I}_{d} \label{v44}
	\end{eqnarray}
	
	By using the above relations between the observables  given by Eq.~(\ref{v43}) and Eq.~(\ref{v44}), one can always construct a set of observables for Alice$_1$ and Alice$_2$. Next, we recall the optimization condition obtained in the SOS method from the Eq.~(\ref{impn3s1}) to find the constraints on Bob's observable as  $M^2_{3,i} \ket{\psi} =0$ \  with $i\in\{1,2,3\}$.
 
In fact, we can find example of such set of observables for qubit system is given by
\begin{eqnarray}\label{obsn3}
 \nonumber
&&A^1_{3,1}\ \ \  =  \ \ \ \sigma_{z}\ \ \  =  \ \ \ A^2_{3,1}\\
\nonumber
 && A^1_{3,2}=\left(\frac{\sqrt{3}}{2}\sigma_{x}+\frac{\sigma_{z}}{2}\right) = A^2_{3,2}\\
  \nonumber
 && A^1_{3,3}=\left(\frac{\sqrt{3}}{2}\sigma_{x}-\frac{\sigma_{z}}{2}\right)=  A^2_{3,3} \\	
\end{eqnarray}
	Note that employing the above-mentioned observables on their respective subsystem, one obtain the quantum optimal value $\left(\beta^2_{3}\right)_{Q}^{opt}=3\sqrt3$ when two maximally entangled two-qubit states are shared between Alice$^1$-Bob and Bob-Alice$^2$.
	
\begin{figure*}[ht]
		\centering
		\includegraphics[width=1.0\linewidth]{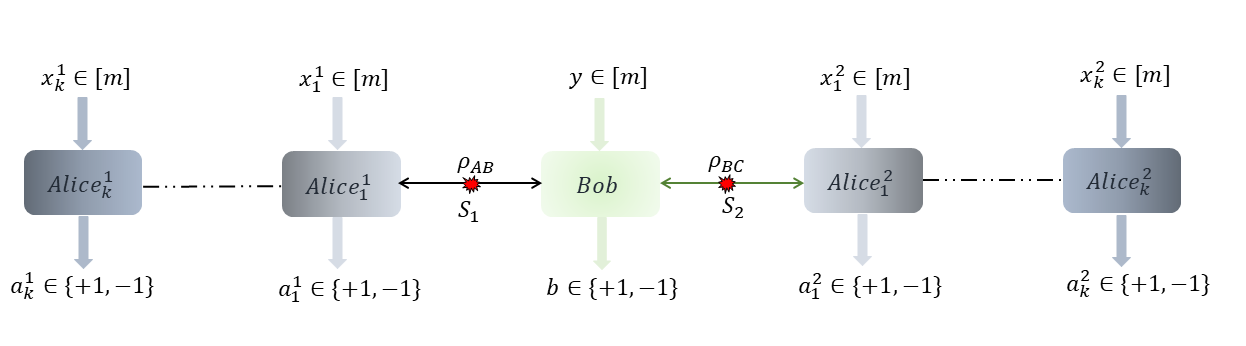}
		\caption{Sequential sharing of the nonlocality in the bilocal network for $m=3$ is depicted. There are $k$ number of sequential Alice$^{1}$ (Alice$^2$) and a central observer Bob. Bob always performs sharp measurements and each Alice$^{1}_{k}$ (Alice$^{2}_{k}$) performs an unsharp measurement and transfers the state to the next sequential observer. The term $k$ is arbitrary here, and our task is to find the maximum value of $k$ up to which nonlocality can be shared in the given scenario.}
		\label{fig2}
	\end{figure*} 
 \section{Sharing of nonlocality in the bilocal network for $m=3$}
Next, we consider the sequential sharing of quantum correlations in the bilocal scenario invoking three inputs for each Alice$^1$, Bob, and Alice$^2$. Alice$^1$ and Alice$^2$ perform unsharp measurements (POVMs) and transfer the state to the next sequential observer respectively as shown in the Fig. 2 Central observer Bob always performs sharp measurements. To perform unsharp measurements one requires positive operator-valued measures (POVMs). These POVMs are the noisy variant of projective measurement. We introduce two varieties of sharing: the symmetric case  in which sequential sharing of nonlocality is demonstrated by both the edge parties, and the asymmetric case in which sequential sharing of nonlocality is explored by one edge party.

\subsection{Symmetric case of sharing nonlocality}
	At first, we introduce sharing in the bilocal scenario i.e., ($n=2$) for the symmetric case by considering three inputs i.e., ($m=3$). We denote Alice$^1$(Alice$^2$) observables as $A^1_1$, $A^1_2$and $A^1_3$ ($A^2_1$, $A^2_2$and $A^2_3$) corresponding to input $x^1_k\in[3]$ ($x^2_k\in[3]$) and Bob's observables as $B_1$, $B_2$ and $B_3$  corresponding to input $y\in[3]$ respectively. As shown in Fig. ({\ref{fig2}}), multiple independent Alice$^1s$ (say, Alice$^1_k$) and Alice$^2s$ (say, Alice$^2_k$) perform the unsharp measurement, where $k$ is arbitrary. Bob always performs sharp measurement. The first Alice$^1$ i.e., (Alice$^1_1$) performs the unsharp measurement on subsystem shared between Alice$^1$ and Bob and transfer the state to the second sequential Alice$^1$ i.e., (Alice$^1_2$) who does the same. The same sequential measurement procedure is adopted by Alice$^2$. We find that, at most, two Alice$^1$ and two Alice$^2$ can share nonlocal correlation in the symmetric scenario.

    We denote the POVMs of Alice$^{1}_{k}$ and Alice$^{2}_{k}$ as $E_{x_1x_2}^{a_1a_2} = E_{x_1}^{a_1} \otimes(\mathbb{I} \otimes \mathbb{I}) \otimes E_{x_2}^{a_2}$ with $\sum E_{x_1x_2}^{a_1a_2}=\mathbb{I}$ and Kraus operators   $ M_{x_1x_2}^{a_1a_2} = M_{x_1}^{a_1} \otimes (\mathbb{I} \otimes \mathbb{I}) \otimes M_{x_2}^{a_2}$, such that $E_{x_1x_2}^{a_1a_2}= \left(M_{x_1x_2}^{a_1a_2}\right)^{\dagger} M_{x_1x_2}^{a_1a_2} $. Here, the inputs $x_1,x_2\in[3]$ and the outputs $a_1,a_2\in \{+1,-1\}$. We take the unbiased POVMs for $k^{th}$ Alice as 
	\begin{eqnarray}
		E_{x_1/x_2}^{\pm}&=&\left( \dfrac{1\pm \lambda_{2,k}}{2} \Pi ^{+}_{x_1/x_2} + \dfrac{1 \mp \lambda_{2,k}}{2} \Pi ^{-}_{x_1/x_2} \right)
	\end{eqnarray}
	where $\Pi ^{\pm}_{x_1/x_2}$ are the projectors satisfying $\Pi ^{+}_{x_1/x_2}+\Pi ^{-}_{x_1/x_2}=\mathbb{I}$, with $A_{x_1}=\Pi ^{+}_{x_1}-\Pi ^{-}_{x_1}$ and $A_{x_2}=\Pi ^{+}_{x_2}-\Pi ^{-}_{x_2}$. Also $\lambda_{2,k}\in [0,1]$.  Consequently, the measurement operators are given by
	\begin{eqnarray}
		\label{kraus}
		M_{x_1/x_2}^{\pm}&=& \sqrt{\dfrac{1\pm \lambda_{2,k}}{2}} \Pi ^{+}_{x_1/x_2} + \sqrt{\dfrac{1 \mp \lambda_{2,k}}{2}} \Pi ^{-}_{x_1/x_2}
	\end{eqnarray}	
	Later we will take $\lambda^{1}_{2,k}\in [0,1]$ is the unsharpness parameter of Alice$^{1}_{k}$. Here subscript $2$ in $\lambda^{1}_{2,k}$ denotes bilocality scenario, i.e., $n=2$, Similar POVMs for $k^{th}$  Alice$^{2}_{k}$ is considered with a different unsharpness parameter $\lambda^{2}_{2,k}\in [0,1]$.
	
 Following the unsharp measurements of the first sequences (i.e., Alice$^{1}_{1}$ and Alice$^{2}_{1}$) the  post-measurement state that is transmited to next sequences of edge observers (Alice$^{1}_{2}$ and Alice$^{2}_{2}$) is given by 
	\begin{eqnarray}
		\label{reduced}
		\rho_{A_1BA_2}^{2}= \dfrac{1}{9} \sum_{\substack{a_1,a_2=+,- \\ x_1,x_2 = 1,2,3}}^{} \left(M_{x_1x_2}^{a_1a_2}\right)^\dagger \rho^1_{A_1BA_2}  \left(M_{x_1x_2}^{a_1a_2}\right)
	\end{eqnarray}
	The final joint state shared between Alice$^{1}_{k}$, Alice$^{2}_{k}$ and BOb after repeating above process for $(k-1)^{th}$ times as 
	\begin{eqnarray}
		\label{rhok}
		\rho_{A_1BA_2}^{k}= \dfrac{1}{9} \sum_{\substack{a_1,a_2=+,-\\ x_1,x_2 = 1,2,3}}^{} \left(M_{x_1x_2}^{a_1a_2}\right)^\dagger \rho^{k-1}_{A_1BA_2}  \left(M_{x_1x_2}^{a_1a_2}\right)
	\end{eqnarray}
	For our purpose, we write the Kraus operator by using Eq. (\ref{kraus}) as 
  \begin{widetext}
	\begin{eqnarray}
		\label{ek}
		\nonumber
		M^{++}_{x_1x_2} = \left( \alpha^1_{k^+} \mathbb{I} + \alpha^1_{k^-} {A_{x_1}}\right)\otimes (\mathbb{I} \otimes \mathbb{I}) \otimes \left ( \alpha^2_{k^+} \mathbb{I} + \alpha^2_{k^-} {A_{x_2}}\right);\ \ \
 		M^{+-}_{x_1x_2}= \left( \alpha^1_{k^+} \mathbb{I} + \alpha^1_{k^-} {A_{x_1}}\right)\otimes (\mathbb{I} \otimes \mathbb{I}) \otimes \left ( \alpha^2_{k^+} \mathbb{I} - \alpha^2_{k^-} {A_{x_2}}\right) \\
		M^{-+}_{x_1x_2}= \left( \alpha^1_{k^+} \mathbb{I} - \alpha^1_{k^-} {A_{x_1}}\right)\otimes (\mathbb{I} \otimes \mathbb{I}) \otimes \left ( \alpha^2_{k^+} \mathbb{I} + \alpha^2_{k^-} {A_{x_2}}\right);\ \ \
		M^{--}_{x_1x_2}= \left( \alpha^1_{k^+} \mathbb{I} - \alpha^1_{k^-} {A_{x_1}}\right)\otimes (\mathbb{I} \otimes \mathbb{I}) \otimes \left ( \alpha^2_{k^+} \mathbb{I} - \alpha^2_{k^-} {A_{x_2}}\right) 
	\end{eqnarray}
	where,

	\begin{eqnarray}
 \label{alphak}
		\nonumber
		\alpha^1_{k^{\pm}}=\dfrac{1}{2\sqrt{2}}\left( \sqrt{1+\lambda^{1}_{2,k}} \pm \sqrt{1-\lambda^{1}_{2,k}} \right);\ \ \ \ \
		\alpha^2_{k^{\pm}}=\dfrac{1}{2\sqrt{2}}\left( \sqrt{1+\lambda^{2}_{2,k}} \pm \sqrt{1-\lambda^{2}_{2,k}} \right)
	\end{eqnarray}
	The state $\rho_{A_1BA_2}^{k}$ as defined in Eq. (\ref{rhok}) can be explicitly written by using Eq. (\ref{ek}) as
	\begin{eqnarray}
		\label{rhok1}
  \nonumber
		\rho_{A_1BA_2}^{k}&=&  4\left(\alpha^1_{k^+} \alpha^2_{k^+} \right)^2 \rho_{ABC}^{k-1} 
  +\frac{4}{3}\left(\alpha^1_{k^+}   \alpha^2_{k^-}\right)^2 \left( \sum_{x_2=1}^{3}(\mathbb{I} \otimes \mathbb{I} \otimes \mathbb{I} \otimes A^2_{x_2}) \rho_{ABC}^{k-1} (\mathbb{I} \otimes \mathbb{I} \otimes \mathbb{I} \otimes A^2_{x_2}) \right) \\
  \nonumber
		&+& \frac{4}{3}\left( {\alpha^1_{k^-}   \alpha^2_{k^+}} \right)^2\left( \sum_{x=1}^{3}  ( A^1_{x_1} \otimes \mathbb{I} \otimes \mathbb{I} \otimes \mathbb{I}) \rho_{ABC}^{k-1} ( A^1_{x_1} \otimes \mathbb{I} \otimes \mathbb{I}\otimes \mathbb{I}) \right)
		+\frac{4}{9}\left( \sum_{ \substack{x_1,x_2=1}}^{3} \left( \alpha^1_{k^-}   \alpha^2_{k^-}\right)^2 ( A^1_{x_1} \otimes \mathbb{I} \otimes \mathbb{I} \otimes A^2_{x_2}) \rho_{ABC}^{k-1} ( A^1_{x_1} \otimes \mathbb{I} \otimes \mathbb{I} \otimes A^2_{x_2}) \right)\\
	\end{eqnarray}
  \end{widetext}
	Using Eq. (\ref{rhok1}), the combination of correlations $(J^2  _{3,1})_k$, $(J^2  _{3,2})_k$ and $(J^2  _{3,3})_k$  in Eq.(\ref{m=3}) for $k^{th}$ Alice$^{1}$ and $k^{th}$ Alice$^{2}$ can be derived as
	\begin{equation}
		\label{I22}
		(J^2  _{3,i})_k = \dfrac{\lambda^{1}_{2,k}\cdot\lambda^{2}_{2,k} }{4^{k-1}} \left[ \prod_{j=1}^{k-1} \left(1+\sqrt{1-\left( \lambda^{1}_{2,j}\right)^2} \right) \left(1+\sqrt{1-\left(\lambda^{2}_{2,j} \right)^2}\right)\right] J^2  _{3,i}
		 \end{equation}
   where $i\in [3]$. The considition to violate the bilocality inequality  in Eq. (\ref{beta23i}) by any $k^{th}$ sequential Alice$^{1}$(Alice$^{1}_{k}$) and Alice$^{2}$(Alice$^{2}_{k}$) is 
	\begin{equation}
		\label{Ik22}
		(\beta^2_{3})^{k}_{Q}=\sqrt{|(J^2  _{3,1})_k|}+\sqrt{|(J^2  _{3,2})_k|} +\sqrt{|(J^2  _{3,3})_k|}>4   
	\end{equation}
	Putting the values of Eq.(\ref{I22}) in Eq.(\ref{Ik22}), we obtain that for the violation of bilocality inequality for the first sequence of edge observers Alice$^{1}_{1}$ and Alice$^{2}_{1}$, the condition 
	\begin{equation}
		\label{25}
		\sqrt{\lambda^{1}_{2,1} \lambda^{2}_{2,1}} \hspace{1mm}(\beta^2_{3})^{opt}_{Q} > 4
	\end{equation}
	has to be satisfied. Here $\lambda^{1}_{2,1}$ and $\lambda^{2}_{2,1}$ are the unsharpness parameter for Alice$^{1}_{1}$ and Alice$^{2}_{1}$.
	
	We define  $\left( \lambda^{1}_{2,1}\right)^{\ast}$ and $\left( \lambda^{2}_{2,1}\right)^{\ast}$ as the critical values of unsharpness parameters which are just enough to violate the bilocality inequality. If the value of  unsharpness parameters are less than $\left( \lambda^{1}_{2,k}\right)^{\ast}$ and $\left( \lambda^{2}_{2,k}\right)^{\ast}$, it will not provide the quantum violation of bilocality inequality. From Eq. (\ref{25}),  we find the critical values of unsharpness parameters of Alice$^{1}_{1}$ and Alice$^{2}_{1}$  which are  $\left( \lambda^{1}_{2,1}\right)^{\ast}= \left( \lambda^{2}_{2,1}\right)^{\ast}=4/3\sqrt{3}\approx 0.77$ respectively. Further, we calculate the upper bound on $\lambda^{1}_{2,2}$ and $\lambda^{2}_{2,2}$ by using above  critical values. For the violation of bilocality inequality in Eq. (\ref{beta23i}) by second sequences of edge observers Alice$^{1}_{2}$ and Alice$^{2}_{2}$ the condition 
	\begin{equation}
		\left(\frac{\lambda^{1}_{2,2} \lambda^{2}_{2,2}}{4} \left[ \left(1+\sqrt{1- \left( \left( \lambda^{1}_{2,1}\right)^{\ast}\right)^2} \right) \left(1+\sqrt{1- \left( \left( \lambda^{2}_{2,1}\right)^{\ast}\right)^2} \right)\right]\right)^{1/2}>\frac{4}{3\sqrt3}
	\end{equation}
	has to be satisfied. Taking same unsharpness parameter for each edge party, we found that for the violation of bilocality by Alice$^{1}_{2}$ and Alice$^{2}_{2}$  the value of unsharpness parameters need to be $\lambda^{1}_{2,2}= \lambda^{2}_{2,2} \geq 0.93$. Following the same line, we find that the values of unsharpness parameters by the third sequences of edge parties i.e., for Alice$^{1}_{3}$ and Alice$^{2}_{3}$,  for violation of bilocality inequality has to be $\lambda^{1}_{2,3}= \lambda^{2}_{2,3} \geq 1.374$, which are not the allowable values. Thus, at most two sequential Alice$^{1}$ and Alice$^{2}$  can share nonlocality  in the symmetric scenario in bilocal network invlving three inputs ($m=3$). However, the above inference is different if we consider  asymmetric case of sharing the nonlocality.

 \subsection{The asymmetric case of sharing nonlocality in bilocal network}
	Now our aim is to find whether it is possible to share nonlocality by more than two sequential  edge observer in bilocal scenario. We find that, in asymmetric case (when sharing of nonlocality is examined across one edge only ) arbitrary $k$ number Alice$^{1}$s can share nonlocality. Here central observer (Bob) and Alice$^{2}$, both perform sharp measurements. For asymmetric case, the combination of correlations as explained in  Eq. (\ref{m=3}) for $k^{th}$ Alice$^{1}$ can be cauculated as 
	\begin{equation}
		\label{Ias22}
		(J^2  _{3,i})_k =\dfrac{\lambda_{2,k}}{2^{k-1}} \left[ \prod_{j=1}^{k-1} \left(1+\sqrt{1-(\lambda_{2,j})^2} \right)\right] J^2_{3,i}
	\end{equation}
	 where $i\in [3]$. We drop the superscript 1 in $\lambda^1_{2,k}$ here as only one edge party i.e., Alice$^{1}_{1}$ is sharing the nonlocality. The quantum violation of bilocality inequality is obtained when the condition   
	\begin{eqnarray}
 \label{a40}
		\left( \dfrac{\lambda_{2,k}}{2^{k-1}} \left[ \prod_{j=1}^{k-1} \left(1+\sqrt{1-(\lambda_{2,j})^2} \right)\right]\right)^{1/2}(\beta^2_{3})^{opt}_{Q}> 4
	\end{eqnarray}
	is satisfied. We again define a critical value  $(\lambda_{2,k})^{\ast}$ for Alice$^{1}_{k}$ of the unsharpness parameter for the asymmetric case which is just enough to satisfy the condition in Eq. (\ref{a40}). For the first Alice$^{1}_{1}$  the critical value is ($\lambda_{2,1})^{\ast}= 0.59$ which is  considerably lower than the critical value  $4/3\sqrt{3}\approx 0.77$ obtained in the symmetric case. This gives us a hint that there is a possibility of sharing nonlocality by more than two sequential observers. We found at most four sequential  Alice$^{1}$  can share nonlocality in the asymmetric case. The critical values of unsharpness parameter $\lambda_{2,k}$ required for violating bilocality inequality in Eq.  (\ref{beta23i}) are given by $\lambda_{2,1}>0.59$, $\lambda_{2,2}>0.66$, $\lambda_{2,3}>0.75$, $\lambda_{2,4}>0.90$.
	For fifth sequential observer the value of unsharpness parameter for  Alice$^{1}_{5}$ has to be $(\lambda_{2,5})^{\ast}>1.25$ which is not a legitimate  value. 
	
One may ask a natural question whether more sequential observers in the asymmetric case can share nonlocality. We answer to this question affirmatively. By increasing the number of edge parties $n>3$ we show that an unbounded number of sequential observers across one edge can share nonlocality for a suitably large value of $n$.
 
 \section{Sharing of nonlocality in the $n$-local star-network for three-input scenario}
	
We consider a star-network configuration that consists of $n$ independent sources and $n + 1$ parties, including a central party Bob and $n$ edge parties say Alice$^{n}$ as shown in Fig. (\ref{nmlocality}). Each source produces an independent physical system. One edge party receives only one physical system from the respective source with which it is connected to Bob. As Bob is connected with each source, he possesses $n$ number of the physical system. Each party receives three inputs i.e., $m=3$ and the measurements produce binary outputs. The following nonlinear $n$-local inequality for three inputs is written as 
	\begin{equation}
		\label{sn}
		\left(\beta^n_{3}\right)_{n-local}=|J^n_{3,1}|^{1/n}+ |J^n_{3,2}|^{1/n} + |J^n_{3,3}|^{1/n} \leq 4  
	\end{equation}
	
	The optimal quantum value is  $(\beta^n_{3})^{opt}_{Q}=3\sqrt{3}$. Following the same line as shown in sec. (\ref{SECVA}), we can find the optimal value without referring to the dimension of the system. The detailed derivation of the optimal value using the SOS approach is placed in Appendix. \ref{appsos}.
 
	For sharing nonlocality, each independent Alice (say, Alice$^n$) performs the unsharp measurement in sequence according to the input $x^n_k\in[3]$ where $n$ denotes the number of Alice and $k$ denotes the number of sequence of the specific Alice. For example, Alice$^1_k$ denotes measurement will perform by $k^{th}$ sequence of first Alice. Bob always performs sharp measurements. Each Alice$^n$ performs the unsharp measurement on the respective physical subsystem shared between each Alice$^n$ and Bob and transfers the post-measurement state to its second sequential observer respectively who performs the same. 
 
We start with the case of $n=3$, i.e.,  the trilocal scenario featuring three edge parties and a central party, We have to find whether sharing of nonlocality is possible for more number of sequential observers in comparison to the bilocal scenario ($n=2$). Each Alice (Alice$^1$, Alice$^2$, and Alice$^3$ ) performs the unsharp measurement on the respective subsystem and transfers the post-measurement state to its second sequential observers respectively. To demonstrate nonlocality, one needs \begin{equation}
\label{cn}
\left(\lambda^{1}_{3,1}\lambda^{2}_{3,1}\lambda^{3}_{3,1}\right)^{1/3} (\beta^n_{3})^{opt}_{Q} > 4
\end{equation}
where $\lambda^{1}_{3,1} $, $\lambda^{2}_{3,1} $ and $\lambda^{3}_{3,1} $ are the unsharpness parameters for Alice$^{1}_{1}$, Alice$^{2}_{1}$ and Alice$^{3}_{1}$ respectively. \\

From Eq. (\ref{cn}) , we find the critical values of unsharpness parameters of Alice$^{1}_{1}$, Alice$^{2}_{1}$ and Alice$^{3}_{1}$ are $\left(\lambda^{1}_{3,1} \right)^{\ast}=\left(\lambda^{2}_{3,1} \right)^{\ast}=\left(\lambda^{3}_{3,1} \right)^{\ast}=4/3\sqrt{3}\approx 0.77$. Using those critical values we can estimate the upper bound on $\lambda^{1}_{3,2} $, $\lambda^{2}_{3,2} $ and $\lambda^{3}_{3,2} $ for the second sequence of edge observers, i.e.,  Alice$^{1}_{2}$, Alice$^{2}_{2}$ and Alice$^{3}_{2}$  respectively. 

We find that for the violation of trilocality inequality (putting $n=3$ in Eq. (\ref{sn} ) for Alice$^{1}_{2}$, Alice$^{2}_{2}$ and Alice$^{3}_{2}$ the values of unsharpness parameters need to be $\lambda^{1}_{3,2} = \lambda^{2}_{2,2} = \lambda^{3}_{3,2}> 0.93$. For Alice$^{1}_{3}$, Alice$^{2}_{3}$ and Alice$^{3}_{3}$ the values of unsharpness parameters has to be $\lambda^{1}_{3,3} = \lambda^{2}_{3,3} = \lambda^{3}_{3,3}\geq 1.3740$, which are not allowable values. Thus for $n=3$, in the trilocal network we find that at most two sequential observer for each Alice can share nonlocality in symmetrical case. We further find that the results remains the same even we increase the number of edge parties. 

\begin{figure}[ht]
\centering
\includegraphics[width=0.9\linewidth]{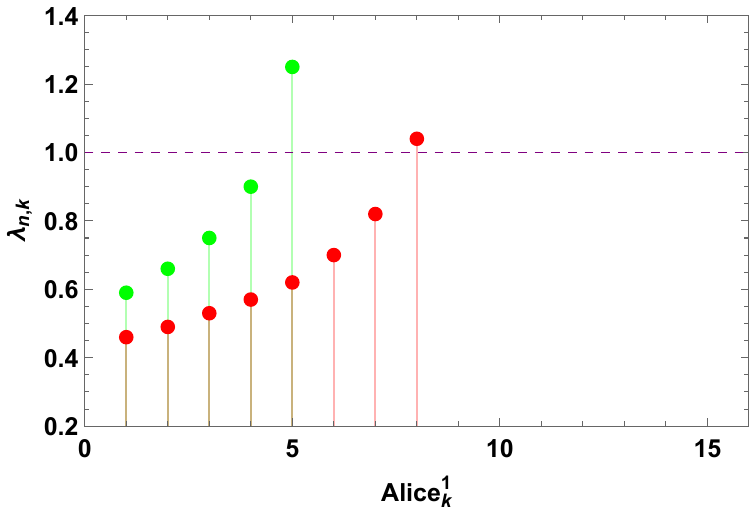}
\caption{critical value of unsharpness parameter $\lambda_{n,k}$ verses number of sequential sharing of nonlocality for $Alice^1$ in asymmetric case required for violating  the $n-locality$ inequality for three-input $(m=3)$ . The green and red dots represent critical values of the unsharpness parameter required for the violation of bilocal inequality $(n=2)$ and the trilocal inequality $(n=3)$ respectively.}
\label{3locality}
\end{figure}

\begin{figure*}[ht]
\centering
\includegraphics[width=0.9\linewidth]{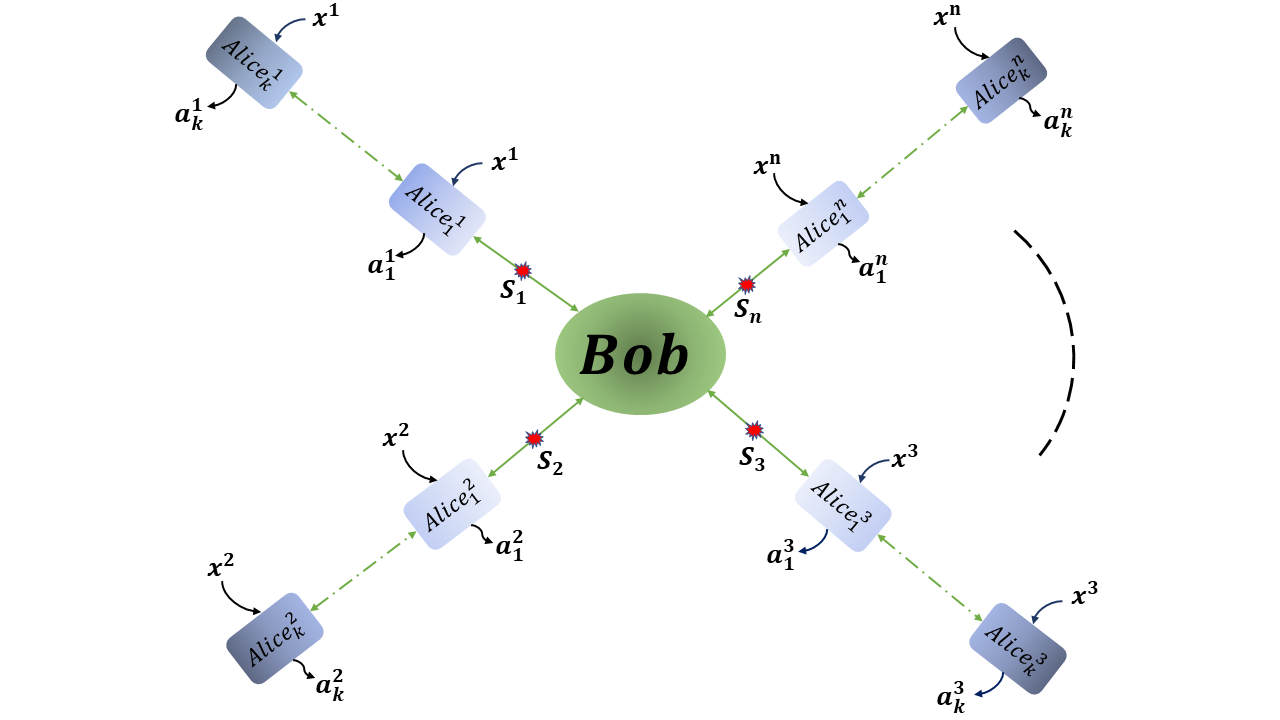}
\caption{This figure consists of $n$ independent sources $S_{n}$ named $n$-local scenario. There are $n$ number of edge parties (Alice$^{n}$) and a central party Bob. Each party receives $m$  inputs, i.e.,  $x^n \in [m]$ and  $y\in [m]$. Each edge party shares a physical system with the central party Bob. Each edge party Alice$_{k}^{n}$ performs unsharp measurements and transfers the state to the next sequential edge observer.}
\label{nmlocality}
\end{figure*}

We find that,  more sequential observers across one edge can share nonlocality for trilocal scenario $(n=3)$ compared to bilocal scenario $(n=2)$. In Fig. (\ref{3locality}), we show the critical value of unsharpness parameter  $\lambda^n_{k}$  and $k^{th}$ sequence of Alice$^1$. In the trilocal scenario, maximum seven $(k=7)$ sequential observer for one edge party can share nonlocality. Here note that except one edge party (named Alice$^1$) all other party will perform sharp measurement.

  Next, we extend the above result for arbitrary $n$ for the asymmetric case. We can intuitively expect that there will be a rise in the number of sequential observer across one edge party (the only party who perform unsharp measurement) who can share nonlocality with the number of an increasing number of $n$. As $n$ is unbounded, there will be an unbounded number of sequential observers across one edge who can share nonlocality in $n$-locality scenario. Assuming all the edge parties Alice$^{n}$ performs sharp measurement except Alice$^{1}$, the condition for obtaining the violation $n$-locality inequality in Eq. (\ref{sn}) for $k^{th}$ sequential observer across one edge is given by
	\begin{eqnarray}
		\label{2nlocal}
		\left(\dfrac{\lambda_{n,k}}{2^{k-1}} \left[ \prod_{j=1}^{k-1} \left(1+\sqrt{1-(\lambda_{n,j})^2} \right)\right]\right)^{1/n}(\beta^n_{3})^{opt}_{Q} > 4
	\end{eqnarray}
	where $\lambda_{n,j}$ is defined the unsharpness parameter for $j^{th}$ sequence of Alice$^{1}$. By using above inequality, we find  the critical value of unsharpness parameter for the first sequential edge observer Alice$^{1}_{1}$ which is $(\lambda_{n,1})^{\ast}= \bigg(\frac{4}{3\sqrt{3}}\bigg)^{n}\equiv \alpha$

Writing $(k-1)^{th}$ sequence with the help of  Eq. (\ref{2nlocal}) and dividing, we get a general condition on unsharpness parameter of $k^{th}$ observer of Alice$^{1}$ (Alice$^{1}_{k}$) for violating the $n$-locality inequality as
	\begin{equation}
		\label{unscdn}
		\lambda_{n,k}\geq \dfrac{2\lambda_{n,k-1}}{1+\sqrt{1-(\lambda_{n,k-1})^2}}
	\end{equation}
	where $\lambda_{n,k}$ and $\lambda_{n,k-1}$ are the critical values of the unsharpness parameter of the  $k^{th}$ and $(k-1)^{th}$ sequential edge observer of Alice$^{1}$. We observe that $1+\sqrt{1-(\lambda_{n,k-1})^2} \geq 2\sqrt{1-(\lambda_{n,k-1})^2}$.  Using this in Eq. (\ref{unscdn}), one can demand that the lower bound on $\lambda_{n,k}$ requires
	\begin{equation}
		\label{unsharp1}
		\lambda_{n,k}\geq \frac{\lambda_{n,k-1}}{\sqrt{1-(\lambda_{n,k-1})^2}} 
	\end{equation}
	Eq. (\ref{unsharp1}) is obtained by using the legitimate limit of the unsharpness parameter, and for small $n$, it overestimates the lower bound of the unsharpness parameter.  Using Eq. (\ref{unsharp1}) and the critical value of Alice$^1$, we can calculate the lower bound  for second sequential observer Alice$^{1}_{2}$ as $\lambda_{n,2}\geq {\alpha}/{\sqrt{1-\alpha^2}}$. Following with the same line, for third sequential observer of Alice$^1$(Alice$^1_3$) we get  $\lambda_{n,3}\geq {\alpha}/{\sqrt{1-2\alpha^2}} $. Then the critical value of unsharpness parameter for $k^{th}$ sequential observer of  Alice$^1$(Alice$^1_k$) is 
	\begin{equation}
		\lambda_{n,k}\geq {\alpha}/{\sqrt{1-(k-1)\alpha^2}}
	\end{equation}
	has to be satisfied. 
	Let we assume the $k^{th}$ measurement to be sharp so that $\lambda_{n,k}=1$ and putting the value of $\alpha$, we then have
	\begin{equation}
		\label{3n}
		k\leq\bigg(\frac{3\sqrt{3}}{4}\bigg)^{2n}
	\end{equation}
 which in turn gives
 \begin{equation}
		\label{log3n}
	n(k)\geq\frac{\log k}{2\log\bigg(\frac{3\sqrt3}{4}\bigg)}	
	\end{equation}
 If one wants to demonstrate the sharing of nonlocality by arbitrary  $k$ sequential observers across one edge party, this requires a suitable value of $n(k)$. We thus demonstrate that an unbounded number of sequential observers across one edge  can share nonlocality in the asymmetric scenario of the star network, as $n$ is unbounded.

 \section{Sharing of nonlocality in the $n$-local star-network for arbitrary-input scenario}
 This scenario features $n$ number of edge parties Alice$^n$ and a central party Bob. For each party, there is an arbitrary $m$ number of measurements, and each measurement has binary outcomes. Each edge party performs unsharp measurement and transfers the post-measurement state to its respective sequential observer. As usual, Bob will always perform the unsharp measurement. Our goal is to find the maximum value of $k$ up to which nonlocality is shared in $n$-locality scenario for the symmetric case of sharing of nonlocality in star-network. The suitable linear combinations of correlations for $m$ inputs and $n$ independent sources are defined as 
	\begin{equation}
		\label{nbell}
		J^{n}_{m,i}=\bigg(\prod_{l=1}^{n}(A_i^l+A_{i+1}^l)B_i\bigg)
	\end{equation}
 where $A_{m+1}^l= -A_1^l$. and $i\in[m]$. We propose the generalized $n$-locality inequality for arbitrary $m$ input scenario is given by
	\begin{equation}
		\label{copt}
		\beta^{n}_{m}=\sum\limits_{i=1}^{m}|J^{n}_{m,i}|^{\frac{1}{n}}\leq (2m-2)
	\end{equation} 
	The optimal quantum value of $(\beta^{n}_{m})_{Q}$ is 
	\begin{equation}
		\label{qopt}
		({\beta^{n}_{m}})_{Q}^{opt}=2m\hspace{1mm} cos(\pi/2m)
	\end{equation}
 which was derived in \cite{munshi2022}. We have also provided a derivation of it in Appendix A.	For any arbitrary $m$, The optimal quantum value in Eq. (\ref{qopt}) is larger than the  $n$-local bound in Eq. (\ref{copt}).
 	
Before going to arbitrary $m$ input $n$-locality scenario, we give a brief overview of sequential sharing of the nonlocality in the bilocal network $n=2$ for $m=4$. In this scenario, the nonlinear bilocal inequality can be written as
\begin{equation}
\label{phd1}
\beta^2_4 = \sum_{i=1}^{4}\sqrt{|J^2_{4,i}|}\leq 6
\end{equation}
Following the sharing scheme discussed earlier the post-measurement state for Alice$^{1}_{k}$, Alice$^{2}_{k}$ and Bob can be written as 
\begin{widetext}
\begin{eqnarray}
\label{phd2}
\nonumber
\rho_{A_1BA_2}^{k}&=&  4\left(\alpha^1_{k^+} \alpha^2_{k^+} \right)^2 \rho_{A_1BA_2}^{k-1}
+ \left(\alpha^1_{k^+}   \alpha^2_{k^-}\right)^2 \left( \sum_{z=1}^{4}(\mathbb{I} \otimes \mathbb{I} \otimes \mathbb{I} \otimes C_{z}) \rho_{A_1BA_2}^{k-1} (\mathbb{I} \otimes \mathbb{I} \otimes \mathbb{I} \otimes C_{z}) \right) \\
\nonumber
&+& \left( {\alpha^1_{k^-}   \alpha^2_{k^+}} \right)^2\left( \sum_{x=1}^{4}  ( A^1_{x} \otimes \mathbb{I} \otimes \mathbb{I} \otimes \mathbb{I}) \rho_{A_1BA_2}^{k-1} ( A^1_{x} \otimes \mathbb{I} \otimes \mathbb{I}\otimes \mathbb{I}) \right) + \frac{1}{4}\left( \sum_{ \substack{x,z=1}}^{4} \left( \alpha^1_{k^-}   \alpha^2_{k^-}\right)^2 ( A^1_{x} \otimes \mathbb{I} \otimes \mathbb{I} \otimes A^2_{z}) \rho_{A_1BA_2}^{k-1} ( A^1_{x} \otimes \mathbb{I} \otimes \mathbb{I} \otimes A^2_{z}) \right)\\
\end{eqnarray}
\end{widetext}
 where $\alpha^1_{k^+}$, $\alpha^2_{k^+}$, $\alpha^1_{k^-}$, and $\alpha^2_{k^-}$ are the same as taken in eqn. (\ref{alphak}). The combination of correlations $(J^2_{4,1})_k$, $(J^2_{4,2})_k$, $(J^2_{4,3})_k$  and $(J^2_{4,4})_k$  in Eq. (\ref{phd1}) for $k^{th}$ Alice$^{1}$ and $k^{th}$ Alice$^{2}$ can be derived as
 
\begin{equation}
\label{phd3}
(J^2_{4,i})_k=\dfrac{\lambda^{1}_{2,k} \lambda^{2}_{2,k} }{4^{k-1}} \left[ \prod_{j=1}^{k-1} \left(1+\sqrt{1-\left( \lambda^{1}_{2,j}\right)^2} \right) \left(1+\sqrt{1-\left(\lambda^{2}_{2,j} \right)^2}\right)\right] J^2_{4,i} 
\end{equation}

  where $i\in [4]$. This means that in order for any $k^{th}$ sequential Alice$^{1}$(Alice$^{1}_{k}$) and Alice$^{2}$(Alice$^{2}_{k}$) to violate bilocality inequality for $m=4$ is 
	\begin{equation}
		\label{phd4}
		(\beta^2_{4})^{k}_{Q}=\sqrt{|(J^2_{4,1})_k|}+\sqrt{|(J^2_{4,2})_k|} +\sqrt{|(J^2_{4,3})_k|} +\sqrt{|(J^2_{4,4})_k|}>6  
	\end{equation}
	While putting the values of Eq. (\ref{phd3}) in Eq. (\ref{phd4}), we get the inequality  for the first sequence of edge observers Alice$^{1}_{1}$ and Alice$^{2}_{1}$ for the violation of bilocality inequality as
	\begin{equation}
		\label{phd5}
		\sqrt{\lambda^{1}_{2,1} \lambda^{2}_{2,1}} (\beta^4_{2})^{opt}_{Q} > 6
	\end{equation}
	 The term  $\lambda^{1}_{2,1}$ and $\lambda^{2}_{2,1}$ denotes  the unsharpness parameter for Alice$^{1}_{1}$ and Alice$^{2}_{1}$. For sharing of nonlocality, above inequality must be satisfied.
	
	We find the critical values of unsharpness parameter of Alice$^1_1$ and Alice$^2_1$ are $(\lambda^{1}_{2,1})^{\ast}= ( \lambda^{2}_{2,1})^{\ast}=6/(4\sqrt{2+\sqrt{2}})$ $\approx 0.81$.  Using the above critical values of the unsharpness parameters of  Alice$^1_1$ and Alice$^2_1$, we can determine the lower bound on the unsharpness parameters of the next sequential edge observers Alice$^1_2$ and Alice$^2_2$ which are denoted as ($\lambda^{1}_{2,2})^{\ast}$ and $(\lambda^{2}_{2,2})^{\ast}$. Following the same line of calculation as shown earlier, for sharing of nonlocality by the second sequence of edge observers Alice$^1_2$ and Alice$^2_2$ one needs
	\begin{equation}
		\left(\frac{\lambda^{1}_{2,2} \lambda^{2}_{2,2}}{4} \left[ \left(1+\sqrt{1- \left( \left( \lambda^{1}_{2,1}\right)^{\ast}\right)^2} \right) \left(1+\sqrt{1- \left( \left( \lambda^{2}_{2,1}\right)^{\ast}\right)^2} \right)\right]\right)^{1/2}> 0.81
	\end{equation}
 Using $(\lambda^{1}_{2,2})^{\ast}$ and $(\lambda^{2}_{2,2})^{\ast}$  we find the required values of the unsharpness parameters as $\lambda^{1}_{2,2}= \lambda^{2}_{2,2} \geq 1.02$, which is not a legitimate value. This means that the second sequential observers cannot share nonlocality. Then in the symmetrical case for $m=4$, at most one sequential  Alice$^1$ and Alice$^2$ can share nonlocality while for $m=2,3$ the second sequences of edge observers can share nonlocality. As we have checked, the result remains the same for any $m\geq 4$. 
 
We now examine whether it is possible to share nonlocality by more than one sequential edge observer in the bilocal scenario for input $m=4$ in the asymmetric case of sharing. We find that in the asymmetric case (when one of the two edge observers performs unsharp measurement ) arbitrary $k$ number Alice$^{1}$s can share nonlocality. The central observer Bob and Alice$^{2}$, both perform sharp measurements. For the asymmetric case, the combination of correlations as explained in  Eq. (\ref{phd4}) for $k^{th}$ Alice$^{1}$ change as
\begin{eqnarray}
\label{Ias22}
(J^2_{4,i})_k =\dfrac{\lambda_{2,k}}{2^{k-1}} \left[ \prod_{j=1}^{k-1} \left(1+\sqrt{1-(\lambda_{2,j})^2} \right)\right] J^2_{4,i}
\end{eqnarray}
 where $i\in [4]$. As shown above, we drop the superscript 1 in $\lambda_{2,k}$ because  only one edge party i.e., Alice$^{1}_{1}$ is sharing the nonlocality. The inequality obtained by using Eq. (\ref{phd5}) for  quantum violation in bilocal scenario for asyymetric case receiving input $m=4$ is   
	\begin{eqnarray}
 \label{phd10}
		\left( \dfrac{\lambda_{2,k}}{2^{k-1}} \left[ \prod_{j=1}^{k-1} \left(1+\sqrt{1-(\lambda_{2,j})^2} \right)\right]\right)^{1/2}(\beta^2_{4})^{opt}_{Q}> 6
	\end{eqnarray}
  We find the critical value for Alice$^{1}_{1}$ is ($\lambda_{2,1})^{\ast}= 0.66$ which is considerably lower than the value $0.81$ obtained in the symmetric case. We derived the critical values for next three sequences as $(\lambda_{2,1})^{\ast}=0.66$, $(\lambda_{2,2})^{\ast}=0.75$, $(\lambda_{2,3})^{\ast}=0.90$. For fourth sequence the critical value of unsharpness parameter for  Alice$^{1}_{4}$ is $(\lambda_{2,4})^{\ast}=1.3$ which is not a legitimate value. 
 
Again a question arises here whether we can find more sequential observers in the asymmetric case for sharing nonlocality. The answer is affirmative. When we increase the number of edge parties to $n$,  we find an unbounded number of sequential observers along one edge can share nonlocality for a suitable value of $n$.
\begin{figure}[ht]
\centering
\includegraphics[width=0.9\linewidth]{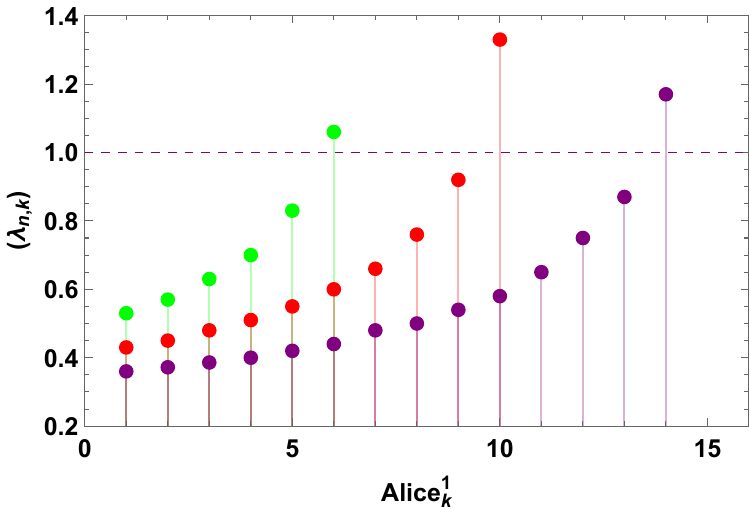}
\caption{The critical values of unsharpness parameter $\lambda_{n,k}$ needed for violating the four-input $(m=4)$  $n-locality$ inequality are shown for $k^{th}$ observer across one edge $Alice^1$. The green, red and purple dots denote the critical values corresponding to $n=3$, $n=4$, and $n=5$ respectively.}
\label{345locality}
\end{figure}
  For concreting the above statement, we calculate the maximum number of sequential observers for which nonlocality can be shared for $n=3$, $n=4$, and $n=5$ taking fixed input $m=4$. We find that at most five, nine, and thirteen (green, red, and purple lines in Fig. (\ref{345locality}).) number of the sequential observer can share nonlocality in the asymmetric case for $n=3$, $n=4$ and $n=5$. We also calculate the critical value of the unsharpness parameter ( green, red and purple dots in Fig. (\ref{345locality})) required for sharing nonlocality in each of the cases. The above-combined data is depicted in  Fig. (\ref{345locality}). Following the increasing number of sequential observers that can share nonlocality with the increase in the value $n$ we can argue that if one wants to demonstrate the sharing for arbitrary $k$ sequential observers, there exist $n(k)$ for which nonlocality can be shared across one edge party. We demonstrate that an unbounded sequential observer can share nonlocality in the asymmetric scenario of the star network, as $n$ is unbounded.  Below we provide an analytical proof of the above statement in $n$-locality scenario for arbitrary $m$ input with allowable approximations. 
  
Assuming sharp measurement perform by all the  parties (Alice$^{n}$ and Bob) except Alice$^{1}$, the condition for obtaining the violation of $n$-locality inequality  for $k^{th}$ sequential observer for $m$ input across one edge (Alice$^{1}$) is given by
	\begin{eqnarray}
		\label{nlocal}
		\left( \dfrac{\lambda_{n,k}}{2^{k-1}} \left[ \prod_{j=1}^{k-1} \left(1+\sqrt{1-\left({\lambda_{n,j}} \right)^2} \right)\right]\right)^{1/n}(\beta^{n}_{m})^{opt}_{Q}\geq 2m-2
	\end{eqnarray}
	where $\lambda_{n,j}$ is the defined as the unsharpness parameter for $j^{th}$ sequence of Alice$^{1}$. By using above 
 
 inequality, we find the critical value of the unsharpness parameter for the first sequential edge observer Alice$^{1}_{1}$ which is
 
 \begin{equation}
     \left(\lambda_{n,1}\right)^{\ast}= \left( \dfrac{2m-2}{2m\hspace{1mm} cos(\pi/2m)}\right)^n\equiv \hspace{2mm}\alpha_m
 \end{equation}
By using Eq. (\ref{nlocal}) the condition for demonstrating nonlocality for $(k-1)^{th}$ Alice$^{1}$ we derive a relation between the respective unsharpness parameters $\lambda_{n,k}$ and $\lambda_{n,k-1}$ of Alice$^{1}_{k}$ and Alice$^{1}_{k-1}$ as 
	\begin{equation}
		\label{unsharpmn}
		\lambda_{n,k}\geq \dfrac{2\lambda_{n,k-1}}{1+\sqrt{1-\left( \lambda_{n,k-1}\right)^{2}}}
	\end{equation}  	
	 By noting $1+\sqrt{1-(\lambda_{n,k-1})^2} \geq 2\sqrt{1-(\lambda_{n,k-1})^2}$ and using it in Eq. (\ref{unsharpmn}), we find the lower bound on $\lambda_{n,k}$ as 
	\begin{equation}
		\label{unsharp}
		\lambda_{n,k}\geq \frac{\lambda_{n,k-1}}{\sqrt{1-\left(\lambda_{n,k-1}\right)^2}} 
	\end{equation}
Note that for small $n$, the lower bound of $\lambda_{n,k}$ is overestimated.  

By using Eq. (\ref{unsharp}) we can calculate the lower bound for the second sequential observer Alice$^{1}_{2}$ as $\lambda_{n,2}\geq {\alpha_{m}}/{\sqrt{1-{\alpha_m}^2}}$ and for the third sequential observer of Alice$^1$(Alice$^1_3$) we get  $\lambda_{n,3}\geq {\alpha_{m}}/{\sqrt{1-2{\alpha_m}^2}} $. We may then write that the unsharpness parameter for $k^{th}$ sequential observer of  Alice$^1$(Alice$^1_k$) has to satisfy
	
	\begin{equation}
		\lambda_{n,k}\geq \frac {\alpha_m}{\sqrt{1-(k-1){\alpha_m}^2}}
	\end{equation}
 If  $k^{th}$ Alice$^1$ performs the sharp measurement so that $\lambda_{n,k}=1$ we get

	\begin{equation}
		\label{nk}
		k\leq\bigg(\frac{2m\hspace{1mm} cos(\pi/2m)}{2m-2}\bigg)^{2n}
	\end{equation}
 where we put the value of $\alpha_{m}$.  From Eq. (\ref{nk}) we can write 
 \begin{eqnarray}
		n(m,k)\geq\frac{\log k}{2\log \bigg(\frac{2m\hspace{1mm} cos(\pi/2m)}{2m-2}\bigg)}
	\end{eqnarray}
	which means that if one wants to share the nonlocality for an arbitrary $k$ number of sequential observers, there exists a suitable $n(m,k)$ as the number of parties in the star-network is unbounded.

\section{SUMMARY AND DISCUSSION}
In this paper, we have studied the sharing of nonlocality in a star network in an arbitrary $m$ input scenario. Such a network features arbitrary $n$ independent sources, $n$ edge parties, and a central party. We note that to derive the optimal quantum violations of $n$-local inequalities for arbitrary $m$ inputs, the dimension of the quantum system is not specified and the inner working of the devices remain uncharacterized.  

We considered two types of sharing.  i) The symmetric case - when the sharing of nonlocality is considered for all the edge parties. ii) The asymmetric case - when the sharing of nonlocality is considered across one edge party. We first consider the bilocal scenario for three-input $m=3$. We showed that at most two sequential observers can share nonlocality in the symmetric case. However, in the asymmetric case, at most four sequential observers can share nonlocality. We extend our study for  $n$-locality scenario by keeping the three inputs for each of the $n$ edge parties and for the central party. We found that again at most two sequential observers can share nonlocality for any value of $n$ in the symmetric case. But for the asymmetric case of sharing an unbounded number of sequential observers can share nonlocality for a sufficiently large value of $n$. 

We further extended our study for $n$-locality scenario while each of the parties receives an arbitrary $m$ number of inputs. We showed that for $m\geq 4$, in the symmetric case of sharing at most one sequential observer can share nonlocality irrespective of the value of $n$. For the asymmetric case, we found that an unbounded number of sequential observers across one edge can share nonlocality for a sufficiently large value of $n$. \\

\section*{Data availability} Data sharing is not applicable to this article as no datasets were generated or analyzed during the current study.

\section*{Conflict of Interest} On behalf of all authors, the corresponding author states that there is no conflict of interest.

\section*{Author contribution statement}
AKP conceived the idea and supervised the work. Both authors contributed to the calculations, and preparation of the manuscript.

\section* {Acnowledgements}
R.K. acknowledges UGC-CSIR NET-JRF
(Fellowship No. 16-6(Dec.2017)/2018(NET/CSIR)] for financial support. A.K.P. acknowledges the support from the project DST/ICPS/QuST/Theme 1/2019/4.

\appendix
\begin{widetext}
 \section{Optimal violation of chain $n$-local inequality optimization through SOS approach}\label{appsos}
Here we provide an analytical proof to find the optimal quantum value of $m$ input $n$-locality scenario through SOS approach, In  Fig. (\ref{nmlocality}), there are $n+1$ parties ($n$ edge parties and a central party), each edge party $A^l$ is connected with the central party through an independent source $S^l$ where $l\in[n]$. All the parties performs $m$ number of dichotomic measurements. We define a suitable linear combination as 
\begin{equation}   
\beta^n_{m}=\sum\limits_{i=1}^{m}| J^n_{m,i}|^{1/n}
\end{equation}
where $J^n_{m,i}=\langle\prod\limits_{l=1}^{n}(A^l_{i}+A^l_{i+1})B_i\rangle$ where  $i\in[m]$ and $A^{l}_{m+1}=-A^l_1$.
By Using the inequality 
\begin{equation}
	\label{Tavakoli}
	\ \forall\ \ z_{k}^{i} \geq 0; \ \ \ \sum\limits_{i=1}^{2^{m-1}}\bigg(\prod\limits_{k=1}^{n}z_{k}^{i}\bigg)^{\frac{1}{n}}\leq \prod \limits_{k=1}^{n}\bigg(\sum\limits_{i=1}^{2^{m-1}}z_{k}^{i}\bigg)^{\frac{1}{n}}
\end{equation}
We can write
\begin{equation}
\beta^n_m\leq \bigg|\prod\limits_{l=1}^{n}\bigg(\sum\limits_{i=1}^{m}(A^l_{i}+A^l_{i+1})B_i\bigg)^{1/n}\bigg|
\end{equation}
With $i\in{m}$ and $l\in{n}$, each obervable is dichotomic. Therefore, we prove that $n$-locality inequality for arbitrary $m$ follows from.  
\begin{equation}
\label{Cmnl}
(\beta^n_m)_{nl}\leq 2m-2
\end{equation}
 First we demonstrate the quantum violation of the inequality in Eq. (\ref{Cmnl})  for $m=3$,  i.e.,  when each  party Alice$^l$ and Bob performs three measurement according to each Alices receiving input  $x^{l}\in[3]$ and and $y\in[3]$. There are $n$ independent sources. Each source produces an independent physical system. One edge party receives only one physical system from the respective source with which it is connected to Bob. As Bob is connected with each source, he possesses $n$ number of physical system. The corresponding inequality is then  given by 
 \begin{equation}
 \label{Xi3}
(\beta^n_{3})_{nl}=|J^n_{3,1}|^{1/n}+|J^n_{3,2}|^{1/n}+|J^n_{3,3}|^{1/n}\leq 4
 \end{equation}
 where $J^n_{3,1}$, $J^n_{3,2}$ and $ J^n_{3,3}$ are suitable linear combinations as follows:
\begin{eqnarray}
J^n_{3,1}&=&\langle\prod\limits_{l=1}^{n}(A_1^l+A_2^l)B_1\rangle;
J^n_{3,2}=\langle\prod\limits_{l=1}^{n}(A_2^l+A_3^l)B_2\rangle;\\
J^n_{3,3}&=&\langle\prod\limits_{l=1}^{n}(A_3^l-A_1^l)B_3\rangle
\nonumber
\end{eqnarray}

  Now,  we  use SOS to optimize $\beta^n_3$ without assuming  the dimension of the system. Following the  method stated earlier in Sec. (\ref{SECVA}), 
  to obtain $(\beta^n_{3})^{opt}_{Q}$, let us consider that $(\beta^n_{3})^{opt}_{Q}\leq\tau^n_{3}$  where $\tau^n_{3}$ is clearly the upper bound of $(\beta^n_{3})_{Q}$.  This is equivalent to showing that there is a positive semidefinite operator $\langle \gamma^n_3\rangle_{Q}\geq 0$ which can be expressed as $\langle \gamma^n_3\rangle_{Q}=-(\beta^n_3)_{Q}+\tau^n_3$.	We define a set of suitable positive operators $M^n_{3,i}$ with $i\in[3]$ which are polynomial functions of   $A^{l}_i$ and  $B_{i}$ where $i\in[3]$ and $l\in[n]$ . We can now write
\begin{equation}
\label{mu3}
\gamma^n_3=\dfrac{(\omega^n_{3,1})^{1/n}}{2}(M^n_{3,1})^{\dagger}M^n_{3,1}+ \dfrac{(\omega^n_{3,2})^{1/n}}{2}(M^n_{3,2})^{\dagger}M^n_{3,2}+\dfrac{(\omega^n_{3,3})^{1/n}}{2}(M^n_{3,3})^{\dagger}M^n_{3,3}\hspace{5mm}
\end{equation}
Here 
 $\omega^n_{3,i}=\prod\limits_{l=1}^{n}(\omega^{A_{l}}_{3,i})$ are suitable positive numbers which  will be specified soon. The optimal quantum value of $(\beta^n_{3})_{Q}$ is obtained if $\langle \gamma^n_3\rangle_{Q}=0$, implying that 
$M^n_{3,i}|\psi\rangle=0$.
Let us  consider  the  positive operators $M^n_{3,i}$ as
\begin{eqnarray}
\label{L3}
\nonumber
|M^n_{3,1}|\psi\rangle|&=&\bigg|\prod\limits_{l=1}^{n}\left(\frac{{A}^{l}_{1}+{A}^{l}_{2}}{(\omega^{A_{l}}_{3,1})}
\right)|\psi\rangle
\bigg|^{1/n} -|B_1|\psi\rangle|^{1/n}\\
|M^n_{3,2}|\psi\rangle|&=&\bigg|\prod\limits_{l=1}^{n}\left(\frac{(A_2^l+A_3^l)}{(\omega^{A_{l}}_{3,2})}
		\right)|\psi\rangle
		\bigg|^{1/n} -|B_2|\psi\rangle|^{1/n}\\
		\nonumber
|M^n_{3,3}|\psi\rangle|&=&\bigg|\prod\limits_{l=1}^{n}\left(\frac{(A_3^l-A_1^l)}{(\omega^{A_{l}}_{3,3})}
		\right)|\psi\rangle
		\bigg|^{1/n} -|B_3|\psi\rangle|^{1/n}
\end{eqnarray}
	where $(\omega^{A_{l}}_{3,1})=||({A}^{l}_{1}+{A}^{l}_{2})|\psi\rangle||_{2}=\sqrt{2+\langle\{A^{l}_{1},A^{l}_{2}\}\rangle}$ and similarly for other $(\omega^{A_l}_{3,i})$s where $l\in[n]$. For notational convenience, we write $|\psi\rangle_{A_1A_2\ldots A_n B}=|\psi\rangle$.       
	Putting Eq. (\ref{L3}) in Eq. (\ref{mu3}), we get 	$\langle\gamma^n_{3}\rangle_{Q}=-(\beta^n_{3})         +(\omega^n_{3,1})^{1/n}+(\omega^n_{3,2})^{1/n}+(\omega^n_{3,3})^{1/n}$. 	Since $\langle\gamma^n_{3}\rangle_{Q}\geq 0$, we have 
\begin{equation}
 \label{C3Qopt1}
(\beta^n_{3})_{Q}^{opt}= max\left((\omega^n_{3,1})^{1/n}+(\omega^n_{3,2})^{1/n}+(\omega^n_{3,3})^{1/n}\right)
\end{equation}
	Using the inequality  Eq. (\ref{Tavakoli})
	we can write
\begin{equation}
\label{C3Qopt}
(\beta^n_{3})_{Q}^{opt}\leq max\bigg[\prod\limits_{l=1}^{n}\bigg((\omega^{A_l}_{3,1})+(\omega^{A_l}_{3,2})+(\omega^{A_l}_{3,3})\bigg)^{1/n}\bigg]\\
\end{equation}
	Now, if we consider the entangled state shared between each Alice's and Bob, and say Bob performs measurement $B_i,i\in[3]$ on his part of subsystem  then the  relevant chained Bell inequality $(\beta^n_{3})_k\leq 4$  can be optimized using the similar SOS  approach as stated above. Hence, the optimal quantum value of chained Bell expression for all $l\in[n]$ is
\begin{equation}
\label{B3kQopt}[(\mathcal{C}^n_{3})_l]_{Q}^{ opt}=max[(\omega^{A_l}_{3,1})+(\omega^{A_l}_{3,2})_+(\omega^{A_l}_{3,3})]
\end{equation}

From Eq. (\ref{C3Qopt}) and Eq. (\ref{B3kQopt}), we can write
\begin{equation}
\label{I3q}
(\beta^n_{3})^{opt}_{Q}\leq\prod\limits_{l=1}^{n}\left([(\mathcal{C}^n_{3})_{l}]^{opt}_Q\right)^{1/n}
\end{equation}
Hence if for a given $l$ ($l\in[n]$), the state $\rho_{A_lB}$ violates relevant chained Bell inequality and each source $S_l$ shares the state $\rho_{A_lB}$ between Alice$^l$ and Bob ($\forall l\in[n]$), then $\bigotimes \limits_{l=1}^{n}\rho_{A_lB}$ violates $n$-locality inequality in  Eq. 
 (\ref{Xi3}).  	
	
Similarly, if we consider the scenario where each party performs four dichotomic observables ($m=4$) then, the corresponding inequality will be 
\begin{equation}
 \label{I4}
\beta^n_{4}=|J^n_{4,1}|^{1/n}+|J^n_{4,2}|^{1/n}+|J^n_{4,3}|^{1/n}+|J^n_{4,4}|^{1/n}\leq 6
 \end{equation}
 where $J^n_{4,1}$, $J^n_{4,2}$, $J^n_{4,3}$ and $J^n_{4,4}$ are suitable linear combinations as follows
\begin{eqnarray}
J^n_{4,1}&=&\langle\prod\limits_{l=1}^{n}(A_1^l+A_2^l)B_1\rangle, \hspace{2mm}
J^n_{4,2}=\langle\prod\limits_{l=1}^{n}(A_2^l+A_3^l)B_2\rangle\\
\nonumber
J^n_{4,3}&=&\langle\prod\limits_{l=1}^{n}(A_3^l+A_4^l)B_3\rangle, \hspace{2mm}
J^n_{4,4}=\langle\prod\limits_{l=1}^{n}(A_4^1-A_1^1)B_4\rangle
\end{eqnarray}

Along the same line of derivation, we can show
\begin{equation}
	\label{I4q}
	(\beta^n_{4})^{opt}_{Q}\leq\ \prod\limits_{l=1}^{n}\left([(\mathcal{C}^n_{4})_{l}]^{opt}_Q\right)^{1/n}
	\end{equation}

	Following this similar approach, in quantum theory, to obtain $(\beta^n_m)^{opt}_{Q}$, we will use SOS approach again. Let us consider $(\beta^n_m)^{opt}_{Q}\leq \beta^n_m$, where $\beta^n_m$ is the upper bound of $(\beta^n_m)^{opt}_{Q}$.  This is equivalent to showing that there is a positive semidefinite operator $\langle \gamma^n_m\rangle_{Q}\geq 0$ which can be expressed as $\langle \gamma^n_m\rangle_{Q}=-(\beta^n_m)_{Q}+\tau^n_m$.	As earlier, by invoking a set of suitable positive operators $M^n_{m,i}$ which are polynomial functions of   $A^{l}_{i}$, $B_{i}$, $i\in[m], l\in[n])$, we can write
\begin{equation}
\label{mum}
\gamma^n_m=\sum\limits_{i=1}^{m}\dfrac{(\omega^n_{m,i})^{1/n}}{2}(M^n_{m,i})^{\dagger}M^n_{m,i}\hspace{5mm}
\end{equation}
and $\omega^n_{m,i}=\prod\limits_{l=1}^{n}(\omega^{A_{l}}_{m,i})$ are suitable positive numbers. The optimal quantum value of $(\mathcal{C}^n_{m})_{Q}$ is obtained if $\langle \gamma^n_m\rangle_{Q}=0$, implying that 
\begin{align}
\label{Lnmi}
M^n_{m,i}|\psi\rangle=0, \forall i\in[m]
\end{align}
We consider  a set of suitable positive operators $M^n_{m,i}$ as
	
\begin{equation}
\label{Lm}
M^n_{m,i}|\psi\rangle=\bigg|\prod\limits_{l=1}^{n}\left(\frac{{A}^{l}_{i}+{A}^{l}_{i+1}}{(\omega^{A_{l}}_{m,i})}
\right)|\psi\rangle\bigg|^{1/n} -|B_i |\psi\rangle|^{1/n}
\end{equation}
where $(\omega^{A_{l}}_{m,i})=||({A}^{l}_{i}+{A}^{l}_{i+1})|\psi\rangle_{A_lB}||_{2}=\sqrt{2+\langle\{A^{l}_{i},A^{l}_{i+1}\}\rangle}$, for each $l\in[n]$.         
	Putting $M^n_{m,i}|\psi\rangle$ of Eq. (\ref{Lm}) in Eq. (\ref{mum}), we get 	$\langle\gamma^n_{m}\rangle_{Q}=-(\beta^n_{m})         +\sum\limits_{i=1}^{m}(\omega^n_{m,i})$. 	Since $\langle\gamma^n_{m}\rangle_{Q}\geq 0$, we have 
\begin{eqnarray}
(\beta^n_{m})_{Q}^{opt}&=& max\bigg[\sum\limits_{i=1}^{m}(\omega^n_{m,i})\bigg]
\end{eqnarray}
If we consider the system shared by the source $S_l$ between  Alice$_l$ and Bob, and say Bob performs measurement $B_i,i\in[m]$ on his part of the sysrem, then the  relevant chained Bell inequality for Alice$_l$ and Bob is given by 
\begin{equation}
(\mathcal{C}^n_{m})_l=\sum\limits_{i=1}^{m}(A^l_i+A^l_{i+1})B_i\leq 2m-2 
\end{equation}
with $A^l_{m+1}=-A^l_{1}$, where $l\in[n]$.
	To optimize $	[(\mathcal{C}^n_{m})_l]_Q$, using  similar SOS approach as stated earlier, we get 
\begin{equation}
[(\mathcal{C}^n_{m})_l]_{Q}^{ opt}=max\sum\limits_{i=1}^{m}(\omega^{A_{l}}_{m,i})=2m\cos\frac{\pi}{2m}
\end{equation}
Now, using the inequality (\ref{Tavakoli}), we get 

\begin{equation*}
(\beta^n_{m})_{Q}^{opt}\leq\prod\limits_{l=1}^{n}\left(\sum\limits_{i=1}^{m}(\omega^{A_l}_{m,i})\right)^{1/n}
\end{equation*}

	which in turn provides 
	\begin{equation}
	\label{Imq}
	(\beta^n_{m})^{opt}_{Q}\leq\ \prod\limits_{l=1}^{n}\left([(\mathcal{C}^n_{m})_l]^{opt}_Q\right)^{1/n}
	\end{equation}
Hence, if for each $l\in[n]$, the state $\rho_{A_lB}$ violates relevant chained Bell inequality and each source $S_l$ shares the state $\rho_{A_lB}$ between Alice$_l$ and Bob, then $\bigotimes\limits_{l=1}^n\rho_{A_lB}$ violates the $n$-locality inequality in star-network configuration  in Eq.(\ref{Cmnl}). Note again that the whole derivation is irrespective of the dimension of the system.

\end{widetext}

\end{document}